\documentclass[authoryear,preprint]{elsarticle}

\usepackage{hyperref}
\usepackage{amsmath,amssymb,amsfonts}
\usepackage{algorithmic}
\usepackage{graphicx}
\usepackage{textcomp}
\usepackage{bm}

\journal{and will appear in Expert Systems with Applications}

\biboptions{authoryear}

\begin{document}

\begin{frontmatter}

\title{Modeling of Electrical Resistivity of Soil Based on Geotechnical Properties}

\author[mymainaddress]{Bandar Alsharari}
\ead{Bandar2@tvtc.gov.sa}

\author[mysecondaryaddress1]{Andriy Olenko\corref{mycorrespondingauthor}}
\cortext[mycorrespondingauthor]{Corresponding author\\
	Phone: +61-3-94792609}
\ead{A.Olenko@latrobe.edu.au} 

\author[mysecondaryaddress2]{Hossam Abuel-Naga}
\ead{h.aboel-naga@latrobe.edu.au}

\address[mymainaddress]{Tabarjal Technical College Branch, Technical and Vocational Training Corporation, Riyadh,  Saudi Arabia}
\address[mysecondaryaddress1]{Mathematics and Statistics Department, La Trobe University, Melbourne,\\ 3086, Victoria, Australia}
\address[mysecondaryaddress2]{Department of Engineering, La Trobe University, Melbourne,\\ 3086, Victoria, Australia\\[3mm]
\url{https://doi.org/10.1016/j.eswa.2019.112966}\\[-10mm]}

\begin{abstract}
Determining the relationship between the electrical resistivity of soil and its geotechnical properties is an important engineering problem. This study aims to develop  methodology for finding the best model  that can be used to predict the electrical resistivity  of  soil,  based on knowing its geotechnical properties. The research develops several linear models, three non-linear models, and three artificial neural network models (ANN).  These models are applied to the experimental data set comprises 864 observations and five variables. The results  show that there are significant exponential negative relationships between the electrical resistivity of soil and its geotechnical properties. The most accurate prediction values are obtained using the ANN model. The cross-validation analysis confirms the high precision of the selected predictive model. 
This research is the first rigorous systematic analysis and comparison of difference methodologies in ground electrical resistivity studies. It provides practical guidelines and examples of design, development and testing non-linear relationships in engineering intelligent systems and applications.
\end{abstract}


\end{frontmatter}
\noindent{\bf Highlights}\\

\noindent\textbullet\ \ Electrical resistivity of soil is modelled by its geotechnical properties.\\
\textbullet\ \ Several linear, non-linear  and artificial neural networks models are developed.\\
\textbullet\ \ Guidelines in the development and quality analysis of the models are provided.\\
\textbullet\ \ The models demonstrate significant exponential negative relationships.\\
\textbullet\ \ Artificial neural networks show much greater accuracy than other models.

%\linenumbers

\section{Introduction}
\label{sec:introduction}
Investigations of the electrical resistivity of soil are important for various industrial applications, for example, estimation of corrosion of underground facilities, analysis of power distribution grounding systems, safe grounding design, see, for example,  \citep{Ackerman, Zhou, Zastrow,  Datsios, Clavel, Gomes}, and the references therein.

The electrical resistivity of geo-materials has attracted the attention of the researchers since 1912 when Schlumberger used the electrical resistivity measurements to characterize the subsurface ground \citep{Griffiths}. Since then, the need to understand the behavior of electrical resistivity of soils has been extended to different areas such as design of electrical earthing system, monitoring moisture content for agriculture applications, electro-osmosis consolidation and drainage of soils, and assessing the homogeneity of compacted clay liner \citep{Laver, Lim, Rhoades, Abu-Hassanein, Tabbagh, Brillante, Jones, Al Rashid}. 

 In most of these areas the number of measurements can be substantial. Even for a few geotechnical soil properties their composition can often result in cases that are not reported in previous studies or available technical documentation. It requires an automatic computer system that can deal with these issues to determine a level of electrical resistivity to be used in practical decision-making solutions. This article discusses and compares several design approaches in this inference analysis.

For engineering design purposes, as the electrical conductivity of soils is a function of its geotechnical properties (soil mineralogy, particle size distribution, void ratio, pore size distribution, pore connectivity, degree of water saturation, pore water salinity, and temperature), it is important to determine a robust quantitative relationship between the electrical resistivity of soils and its geotechnical properties, see \citep{Kibria}.

In the vast majority of applications determining soil electrical resistivity is done by trial and error. The results are often heavily based on empirical knowledge and can be influenced by subjective perceptions.  There is a need to develop rigorous automatic methodologies allowing to estimate electrical resistivity with more certainty. To achieve this, it is important to evaluate various alternatives approaches.

Several laboratory studies have collected data on electrical resistivity of soils and their physical properties, see, for example, \citep{Edlebeck,  Zhou, Southey, Al Rashid, Alipio, Alipio2,Mokhtari}. However, only a few  electrical resistivity models were used to assess the role of different physical properties. Unfortunately, none of these articles justified their model selection. No rigorous statistical evidence was found that can help to determine the best model. The aim of this study is to develop several statistical models of electrical resistivity and to provide general methodology for their comparison and performance evaluation. This systematic approach helps in selecting the  most powerful predictive~model.

It is worth mentioning that while artificial neural networks and other advanced computational methods were used for modeling  characteristics of materials, see, for example, \citep{Gen, Gusel, Hwang}, this research is the first rigorous systematic analysis and comparison of difference methodologies in ground electrical resistivity studies. This paper provides practical guidelines and examples of modelling and investigating non-linear relationships in engineering applications.

Extensive experimental study was conducted in the Department of Engineering at La Trobe University to investigate the effect of soil minerology, moisture content, water salinity, and dry unit weight on the electrical resistivity measurements. These experimental data were used to develop a linear regression model, an exponential non-linear regression model, multivariate adaptive regression splines (MARS), and the artificial neural network model (ANN) for predicting electrical resistivity of soil. Several other models were investigated, but are not reported in this paper as their performances were not satisfactory. 

While this paper concentrates on the design, development and testing of non-linear models in engineering soil studies the presented methodology can be applied to much wider classes of intelligent systems. In particular, the discussed ANN models can provide the overall trends in rather complex relationships even from the limited data.

All computations, data transformations, visualisations, statistical analysis, and models fitting were performed using R~3.5.0 software, in particular, the packages plyr, nlme, car, hexbin, MASS, earth, nnet, and devtools. The data and R codes used in this article are freely  available in the folder "Research materials" from  \url{https://sites.google.com/site/olenkoandriy/}.

\section{Data}\label{Data}

\begin{figure}[h!]
	\centering
	\includegraphics[width=0.8\linewidth]{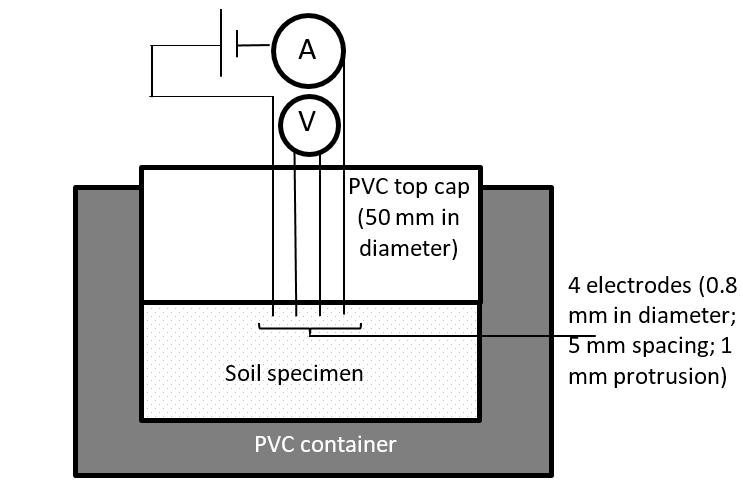}
	\caption{Test apparatus}
	\label{fig:screenshot0011}
\end{figure}

The electrical resistivity measurements were obtained using a soil box shown in Figure~\ref{fig:screenshot0011}. The measurements process was conducted using the 
Wenner four-electrode method, \citep{Wenner, ASTM}. 
A DC test voltage with the magnitude 12V is applied between the outer electrodes,
where the corresponding current, $I,$ and the voltage drop between the
inner electrodes, $V,$ are measured. The resistivity, $ER,$ is then $ER = 2\pi aV/I.$ See more details about the measurement process in  \citep{Al Rashid}. 

The full data set consists of 864 observations with five variables. The variables of interest for this study are: 
\begin{itemize}
	\item Electrical resistivity (ER): This is a continuous variable that measures the electrical resistivity of soil in Ohm-m (the response variable).
	\item Soil water salinity (Mol): This is a continuous variable that measures the amount of NaCl salt in the water in mol unit and has only    four values: 0, 0.5, 1 and 2 Mol.
	\item Soil moisture content (Moist): This is a continuous variable that measures the percentage of gravimetric water content. It is the ratio of weight of water to the weight of solid soil particles. 
	\item 	Dry unit weight (Uw): This is a continuous variable that measures the dry unit weight in  $kN/m^3$.
	\item 	Soil type (ST): This is a factor  (categorical)  variable  that  identifies the soil type. Nine different soil types were test in the conducted experimental study. They comprise different mixtures of Kaolin-Bentonite (K-B) and Kaolin-Sand (K-S) as shown in Table~\ref{ST summary} which also provides the number of observations for each type of soil.  The engineering properties of Kaolin, Bentonite, and Sand used in this study can be found in \citep{Al Rashid}. 
\end{itemize}

Table~\ref{fig:screenshot038}  provides a snapshot of a part of the original data frame for only two soil types. Namely, pure kaolin (100\%K) and kaolin with 10\% of bentonite (90\%K+10\%B).
\begin{table}[h!]
	\caption{Part of original data}
	\label{fig:screenshot038}
	\includegraphics[width=01\linewidth]{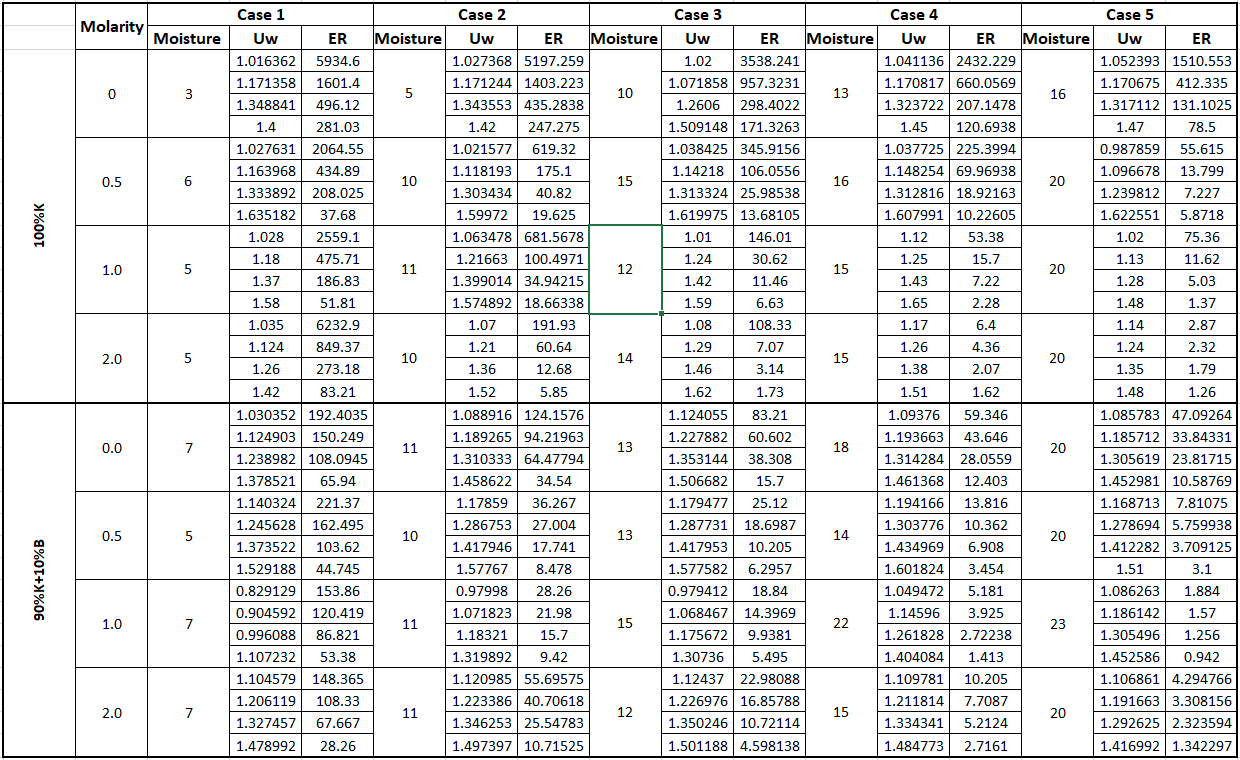}
\end{table}

Table~\ref{Table 1} shows the summary statistics  of all  continuous variables, namely   electrical resistivity, molarity, moisture and unit weight. 
\begin{table}[t!]
	\renewcommand{\arraystretch}{1.3}
	\caption{Summary statistics of the continuous variables} 
	\label{Table 1}
	\centering
	\begin{tabular}{ccccc}
		\hline\\[-1mm]
		&    Mol &    Moist &       Uw &       ER \\[1mm]  
		\hline\\[-2mm]
		Min. :  &   0.000   & 3.00   & 0.7158   &  0.183   \\ 
		1st Qu.: &  0.375   & 10.00   & 1.1570   &   2.890   \\ 
		Median : &  0.750   & 15.00   & 1.3279   &  11.707   \\ 
		Mean   : & 0.875   & 14.93   &1.3321   & 98.203   \\ 
		3rd Qu.: &1.250   & 20.00   & 1.4927   &  46.508   \\ 
		Max.   : &2.000   & 27.00   & 1.9200   &6232.900   \\ 
		NA's   : &        &144      & 144      & 144   \\ 
		\hline 
	\end{tabular}
	
\end{table}

Table~\ref{ST summary} provides the number of observations for each type of soil.
\begin{table}[h!]
	\caption{Sample sizes for each soil type} 
	\label{ST summary}
	\centering
	\begin{tabular}{cc}
		\hline\\[-1mm]
		Soil type		&    n  \\[1mm]
		\hline\\[-2mm]
		
		100\%K           & 92  \\
		90\%K+10\%B      & 92  \\
		80\%K+20\%B      & 96 \\
		70\%K+30\%B      & 104  \\ 
		60\%K+40\%B      & 92   \\
		90\%K+10\%S      & 92 \\
		80\%K+20\%S      & 104  \\   
		70\%K+30\%S      & 96  \\
		60\%K+40\%S      & 96 \\
		\hline
	\end{tabular}
\end{table}

\begin{figure}[h!]
	\centering
	\includegraphics[width=01\linewidth]{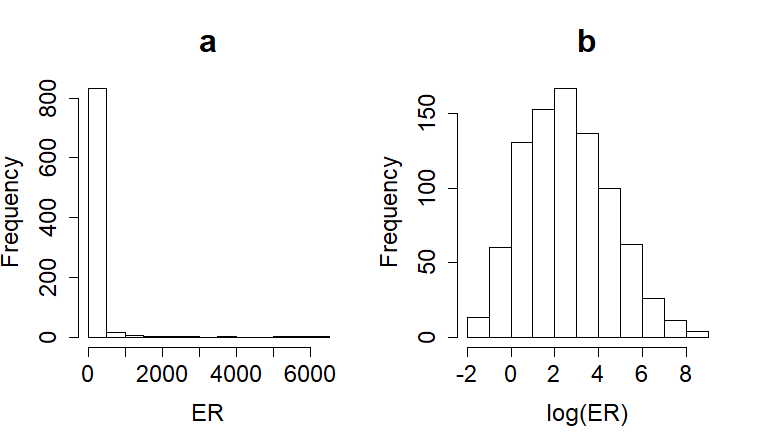}\\[-2mm]
	\caption{Histograms of the electrical resistivity and the log  resistivity.}
	\label{fig:screenshot001}
\end{figure}

A histogram of the electrical resistivity of soil is shown in Figure~\ref{fig:screenshot001}a. This distribution  is highly skewed to the right, indicating  that ER is abnormally distributed (potentially exponentially distributed). However, the histogram of log(ER), which  is shown in Figure~\ref{fig:screenshot001}b,  is  much closer to the normal distribution.   Therefore,  the log  or similar transformations of the ER variable might be appropriate when fitting the linear regression model. 
\vspace{-3mm}

\section{Multiple regression analysis }\label{Multiple regression}
\subsection{Ordinary least squares approach (OLS) }

First,  the function \texttt{lm} of the R software was used to fit the multiple linear regression model to the data set, where the electrical resistivity of soil  is the outcome variable, and  the predictors are  molarity, moisture  and unit weight  in addition to the soil type (where this variable is treated on the base of the  percentage of kaolin that is  mixed with bentonite  and~sand). 

\begin{figure}[h!]
	\centering
	\includegraphics[width=.9\linewidth]{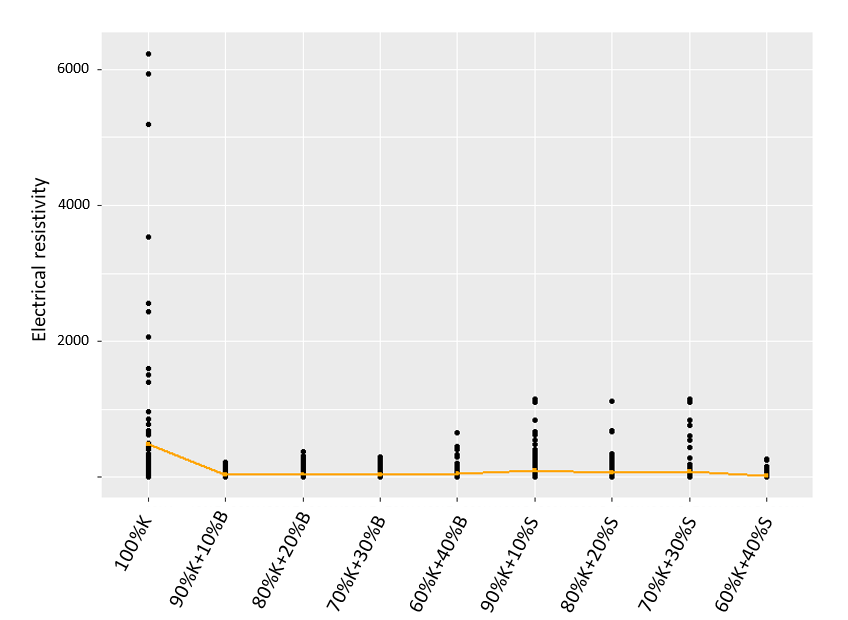}\\[2mm]
	\caption{Scatter plot of ER vs soil type. The line represents the  mean values.}
	\label{fig:Scatter ER vs Soil type}
\end{figure}
Figure~\ref{fig:Scatter ER vs Soil type}  displays  that the electrical resistivity  for 100\%K composition is likely to  yield variation higher than those from other types of soil. There is a  big difference in variability of ER for the soil that consists of pure kaolin and the variability of ER for soil that mixed with either bentonite or sand.   This may  indicate  that treating the soil composition variable on the basis of pure and not pure kaolin  leads to an improvement in our liner model. 

\begin{figure}[h!]
	\centering
	\includegraphics[width=01\linewidth, trim={0cm 0 0 1.5cm},clip]{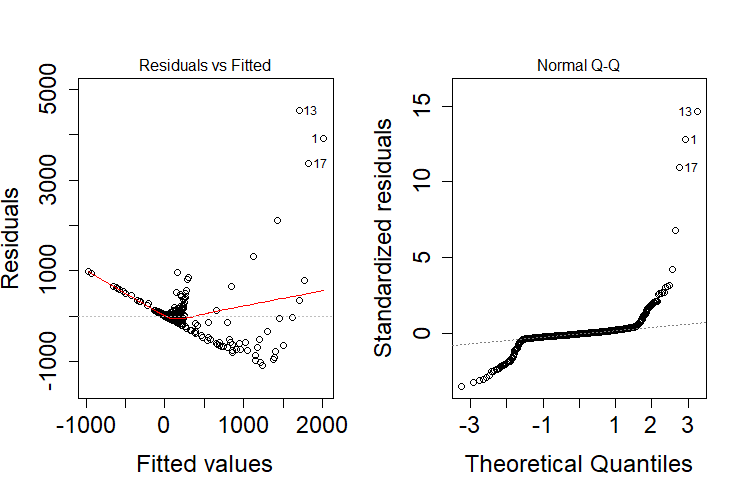}
	\caption{Residuals vs fitted and Normal Q-Q plots for Model~(\ref{model3}).}
	\label{fig:screenshot032}
\end{figure}

Hence, we fit the following three models:
\begin{align}\label{model1}
\widehat{ER}&=1101.2 -643.7 ST.KB -247.3 ST.KS  -59 Mol -14.8 Moist -473.3 Uw \notag\\
&(Adj~R^2= 0.1623)\\
\label{model2}
\widehat{ER}&=1202.5 -373.45 ST.K         -59.7 Mol -14.4 Moist -378.1 Uw \notag\\
&(Adj~R^2= 0.2057)\\
\label{model3}
\widehat{ER}&= 4829.4 -4316.8 ST.K         -44.1 Mol -82.9 Moist -2523.7Uw\notag\\
& + 75 ST.K\times Moist + 2297.5 ST.K\times Uw \notag\\
&(Adj~R^2= 0.4212)
\end{align}
It can be seen that the adjusted $R^2$ for model~(\ref{model1}) is 16.23\%, where  $ST.KB$ and  $ST.KS$ represent  the percentage of bentonite and sand that mixed with kaolin respectively. While in model~(\ref{model2}), these two variables are replaced by $ST.K$ which results in an increase in the adjusted $R^2$  from 16.23\% to 20.57\%.  This means it is  better to treat the soil composition variable  on the basis of whether it is mixed with bentonite or sand, or not. 
Moreover, the adjusted $R^2$ increased from 20.57\% to 42.12\% (Model~(\ref{model3})). Hence, the interaction effect between soil type and Moisture ($ST.K\times Moist$) and interaction effect between soil type and unit weight ($ST.K\times Uw $) variables are significant predictors. Model~(\ref{model3}) explains  42.12\% of the variability in the electrical resistivity of soil  using this data. However, the validity of inference from this model depends on the linear regression assumptions being fulfilled. A violation of the underlying assumptions may lead to an invalid conclusion and an opposite conclusion can be obtained  when using different samples \citep{Montgomery}.  The main assumption is  that the errors are   from a normal distribution ($\varepsilon \sim N(0,\sigma^2)$). So, we check error normality and constant error variance assumptions.  Figure~\ref{fig:screenshot032} illustrates the residuals vs fitted plot and normal Q-Q plot. A normal Q-Q plot raises concerns over the underlying normality. In addition, the residuals vs fitted plot shows that the  constant error variance assumption is violated. There seems to be an increase in  error variance as the fitted values increase. Hence, the detection of such problems undermines the validity of first simple linear models.

\subsection{Box-Cox transformation}\label{S:log Transformation}

The relationship between the dependent   and  independent variables is not a linear relationship. This may be the reason behind  the problems that were found in the previous models in addition to the small value of $R^2$.

The non-normality in the response variable   often causes issues in the OLS technique. A violation of this assumption may lead to a wrong conclusion. Therefore,   this problem needs to be addressed before using the OLS technique. One way to deal with non-normality in the response variable is  using Box-Cox transformation. In the data set, all values of the response variable are  positive ($ER>0$).  According to \citep{Sakia}, for  positive values of the response variable, ($y_i$), the family of the Box-Cox transformation is defined as:
\begin{align*}
	y_i^\lambda=  
	\left\{
	\begin{array}{ll}
		(y_i^\lambda-1)/\lambda,  & \mbox{if } \lambda \ne 0 ,\\
		\log(y_i)   ,              & \mbox{if }  \lambda =  0.
	\end{array}
	\right.
\end{align*}

The parameter  $\lambda$  can be  estimated using the profile likelihood function. The dependent variable was transformed using the Box-Cox transformation. The estimated value of $\lambda$ was very close to~0 ($\lambda=-0.05$). Consequently, the log  of the ER variable might be an appropriate transformation. So, the natural logarithm was applied to the dependent variable and  the models were refitted again as it is shown below.
\begin{align}\label{model11}
	\log(\widehat{ER})&=11.705 -1.26ST.KB  -0.72ST.KS   -0.911Mol  -0.206Moist \notag\\ 
	& -3.802Uw \quad 	(Adj~R^2=0.8265)\\
	\label{model22}
	\log(\widehat{ER})&=12.14 -0.98 ST.K  -0.91 Mol  -0.20Moist  -3.67Uw \notag\\ &(Adj~R^2=0.8449 )\\
	\label{model33}
	\log(\widehat{ER})&= 16.05 -5.23 ST.K  -0.90Mol  -0.246Moist\notag\\
	&  -6.30Uw +0.05ST.K\times Moist +  2.83 ST.K\times Uw  \notag\\ & (Adj~R^2= 0.8541 )  
\end{align}

The R-squared and	adjusted R-squared values significantly  increased  for  these three models which is a substantial improvement of the models. 

\begin{figure}[h!]
	\centering
	\includegraphics[width=01\linewidth]{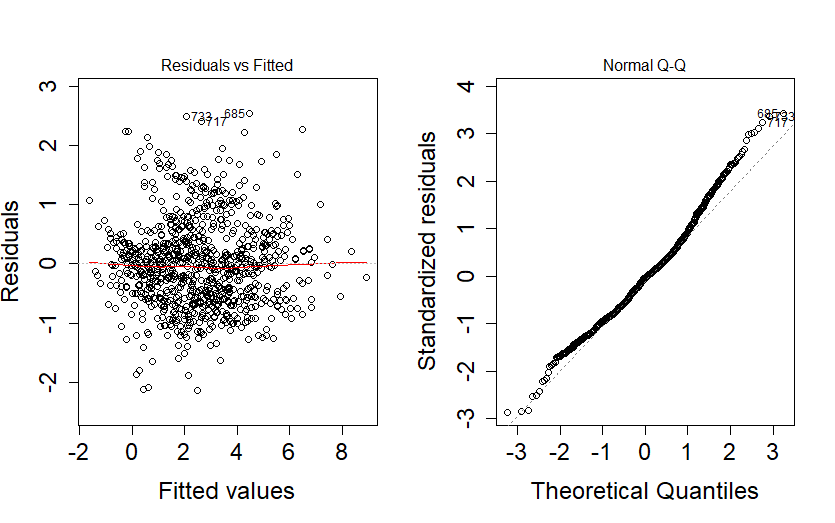}
	\caption{Residuals vs fitted and Normal Q-Q plots for Model~(\ref{model33}).}
	\label{fig:screenshot043}
\end{figure}

To check the validity of the assumptions the  residual vs fitted plot was examined. For   model~(\ref{model33}), the results are shown in Figure~\ref{fig:screenshot043}. The   residual vs fitted plot shows that there is no dependency between residuals and fitted values which indicates that there are no violations to errors normality.
Additionally, the normal Q-Q plot for  model~(\ref{model33}) shows that the points fall close to the 45-degree reference lines with a small deviation at the upper tails. Therefore, the normality is probably a reasonable assumption.

\section{Nonlinear regression model}\label{Nonlinear Model}
Fitting a linear model provides a reasonable first approximation. However, the relationship between the electrical resistivity of soil and its geotechnical properties is not a linear relationship. Therefore, the use of a non-linear  regression model may result in a better fit of the data. Furthermore, using a non-linear model makes the interpretation  of the  estimated parameters clearer.

According to  \citep{Fox}, the general form of non-linear regression models can be written as 
\begin{align*}
	y=E(y|\bm x)+\varepsilon=f(\bm{x}|\bm \theta) +\varepsilon.
\end{align*}

The best estimate of $\bm \theta$  minimizes the residual sum of squares. Unlike linear models, one needs to use an  iterative numerical  procedure to estimate  $\bm \theta$. The non-linear least-squares algorithm may not converge if inappropriate starting values are provided.
Therefore, appropriate starting values need to be selected \citep{Fox} to  estimate  the parameters. 

Due to the  nature of the data (see the discussions in the previous sections), the non-linear model that may fit the data well is
\begin{align}\label{non-linear}
	ER&=\beta_0 +\beta_1\exp(\beta_2ST.KB) \exp(\beta_3ST.KS)\notag\\ 
	&\times\exp(\beta_4Mol) \exp(\beta_5Moist) \exp(\beta_6Uw) +\varepsilon.\hspace{-4mm}
\end{align}

This model was fitted using the \texttt{nls} function of the R computer package \texttt{nls}. The appropriate starting values can be obtained by fitting  a linear model to the data  and taking its coefficients as starting estimates. 
The starting values to estimate $\beta_0$, $\beta_1$, $\beta_2$, $\beta_3$, $\beta_4$, $\beta_5$, $\beta_6$ and  $\beta_7$ in model~(\ref{non-linear}) were obtained from
\begin{align*}
	\log(\widehat{ER}_i)&=11.70548-0.12629 ST.KB_i-0.07152ST.KS_i   \notag\\ 
	&-0.91053 Mol_i  -0.20625Moist_i  -3.80215Uw_i. 
\end{align*}
The  estimates of the coefficients, together with their corresponding standard errors, observed test statistics and  p-values of  Model~(\ref{non-linear}) suggest that the predicted value of ER is given by
\begin{align}\label{predictednon-linear}
	\widehat{ER}&=44.63 +635736041\exp(-5.074ST.KB -4.042ST.KS  \notag\\
	& +0.09561 Mol-0.1968Moist -10.76Uw) 
\end{align}

Unfortunately, the predicted ER using this model cannot be less than the value of the intercept (44.63). However, we have some actual values of ER that are smaller than this value. As a result, this model will fail to accurately  predict electrical resistivity when its actual value is less than 44.63. In addition,  there is a positive relationship between ER and Mol (since the estimate of $\beta_4>0$) which contradicts to the previous models. Therefore, we  refine the functional relationship between the response and the predictors and obtain the  estimated equation
\begin{align}\label{non-linear11}
	\widehat{ER}&= 78.37+318382653 \exp(-48.25ST.KB) \exp(-39.082ST.KS)
	\notag\\
	&   \times\exp(-40.64Moist) \exp(-0.1878Uw)-10.08Mol.
\end{align}

We also refit the model  using  the ST.K variable instead of ST.KB and ST.KS. Namely, we fit the following model
\begin{align}\label{non-linear3}
	ER&=\beta_0 +\beta_1 \exp(\beta_2ST.K) \exp(\beta_4Moist) \exp(\beta_5Uw) +\beta_3Mol +\varepsilon.
\end{align}

The R output below demonstrates that all variables in model~(\ref{non-linear3}) are significant now.
{\footnotesize 
	\begin{verbatim}
	Formula: ER~beta0+beta1*exp(beta2*ST.K+beta4*
	Moist + beta5 * 	Uw) + beta3 * Mol
	Parameters:Estimate Std. Error tvalue  Pr(>|t|)    
	beta0      5.762e+01  1.057e+01   5.454 0.000
	beta1      1.455e+08  6.189e+07   2.352 0.019 
	beta2     -3.602e+00  1.319e-01 -27.313 0.000
	beta3     -3.297e+01  8.896e+00  -3.706 0.000
	beta4     -1.865e-01  7.133e-03 -26.141 0.000
	beta5     -9.330e+00  4.117e-01 -22.665 0.000
	
	Residual standard error: 190.6 on 858 degrees
	of freedom
	Number of iterations to convergence: 17 
	Achieved convergence tolerance: 3.4e-06
	\end{verbatim}
}
Hence, the fitted non-linear model~(\ref{non-linear3}) gives the predicted value of ER  by
\begin{equation}\label{non-linear33}
	\widehat{ER}=  57.62 +145546403\cdot 36.67^{-ST.K}1.20^{-Moist}11275.04^{-Uw}-32.97Mol.
\end{equation}

\begin{figure}[h!]
	\centering
	\includegraphics[width=1\linewidth, height=6.5cm]{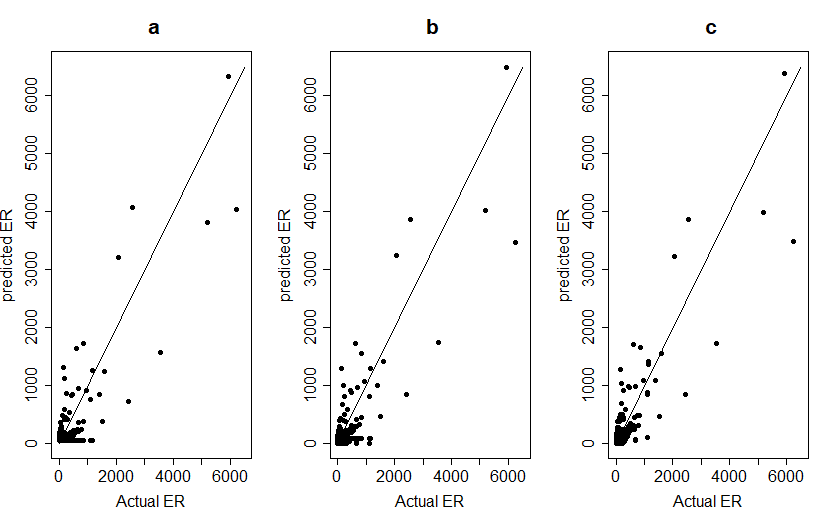}
	\caption{Predicted vs actual values plots for each model.}
	\label{fig:screenshot058}
\end{figure}

Figure~\ref{fig:screenshot058}  shows   the predicted vs actual plots for each non-linear model. Plots a, b and c  are for models~(\ref{predictednon-linear}), (\ref{non-linear11}) and (\ref{non-linear33}) respectively.

We choose the best non-linear model  based on the smallest value for AIC:

{\small 
	\begin{verbatim}
	> AIC(Model.11,Model.13,Model.15)
	df  AIC
	Model.11  8 11654.63
	Model.13  8 11642.82
	Model.15  7 11532.53
	\end{verbatim}
}

The MSE values of models~(\ref{predictednon-linear}), (\ref{non-linear11}) and (\ref{non-linear33}) are  41471.40, 40908.61 and 36089.50 respectively. Hence, the best  of these models  is model~(\ref{non-linear33}) since it has the smallest value of AIC and the smallest value of MSE. 

\begin{figure}[h!]
	\centering
	\includegraphics[width=0.7\linewidth,height=6cm]{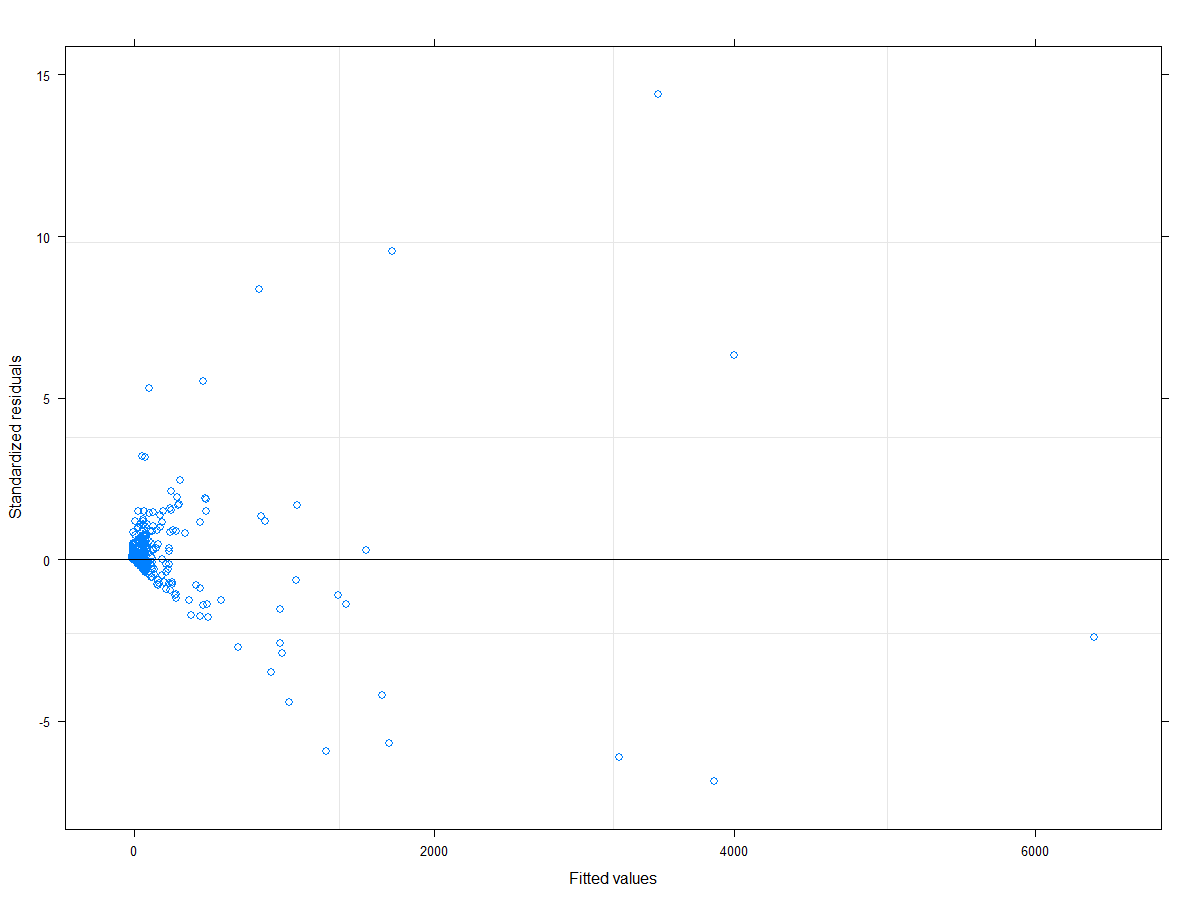}
	\caption{Residuals vs fitted values plot for model\ref{non-linear33}.}
	\label{fig:screenshot042}
\end{figure}

Figure~\ref{fig:screenshot042} shows  the residuals vs fitted values plot for model~(\ref{non-linear33}). 
Overall, there is  concern about the dependency of errors and the magnitude of residuals increases with the increase of fitted values.

\section{Multivariate adaptive regression splines }\label{MARS}
The article \citep{Friedman} introduced multivariate adaptive regression splines (MARS) which is a flexible  regression model that can be used to handle both linear and non-linear relationships. The significant feature of MARS lies in its ability to generate simple and easy-to-interpret models that capture mapping in multi-dimensional data \citep{Zhang}.

The MARS model has the form \citep{Hastie}:
\begin{align*}
	f(X)=\beta_0 +\sum_{m=1}^{M} \beta_m h_m (X),
\end{align*}
where $h_m (X)$  represents a  basis function (hinge function) or product of two or more such functions. Each basis function consists of  the pair $\max(0,x-c)$ and $\max(0,c-x)$ where $c$ is a knot. Each pair is multiplied by a parameter which is estimated by minimizing the residual sum of squares (RSS).

The MARS model divides the data into  regions and fits each region with an appropriate model.

Now, we  refit models~(\ref{model11}), (\ref{model22})  and (\ref{model33}) using the MARS approach. The MARS models were fitted by using the R package "earth". The results of the fits to the ER data are available in the folder "Research materials" from  \url{https://sites.google.com/site/olenkoandriy/}.

\begin{figure}[h!]
	\centering
	\includegraphics[width=01\linewidth,height=7cm]{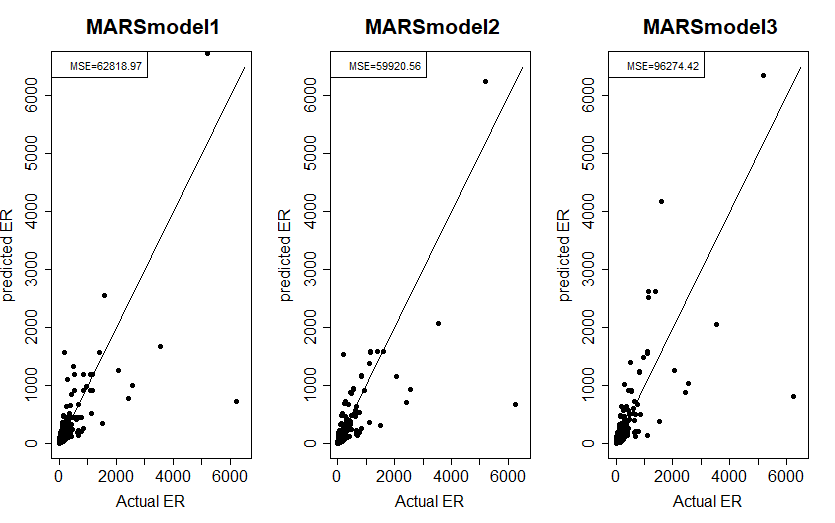}
	\caption{Predicted  ER vs observed ER  for the   MARS models.}
	\label{fig:screenshot052}
\end{figure}

Figure~\ref{fig:screenshot052}  shows the predicted  ER vs the observed ER  for the three MARS models. 

To compare these MARS models and to check their accuracy,  the MSE for each model was calculated. Table~\ref{mseMARS} shows the obtained MSE values.

\begin{table}[h!]
	\caption{MSEs for MARS models} 
	\label{mseMARS}
	\centering
	\begin{tabular}{c|ccc}
		\hline\\[-1mm]
		& MARS1	&  MARS2  & MARS3  \\[1mm]
		\hline\\[-1mm]
		MSE & 62818.97 & 59920.56  & 96274.42 \\
		\hline
	\end{tabular}
\end{table}

Based on Figure~\ref{fig:screenshot052} and the MSE values, the second MARS model gives the best fit of  the three MARS models. Its estimated equation is 
\begin{align*}
	\widehat{ER} = \exp[&4.958 -   0.8739  ST.K  +   2.808  \max(0,  0.5 -  Mol)    -    0.9058 \notag\\ 
	& \times\max(0,   Mol -   0.5)+     0.739  \max(0,   Mol -     1)  -    0.2447  \notag\\ 
	& \times \max(0, Moist -     5) +   0.07649  \max(0, Moist -    15)  \notag \\
	&+     105.9  \max(0,    Uw - 1.005)  -       119  \max(0,    Uw - 1.014) \notag\\
	& +     12.28  \max(0,    Uw - 1.099)  +     2.504 \max(0, 1.663 -    Uw) ]. 
\end{align*}

\begin{figure}[h!]
	\centering
	\includegraphics[width=01\linewidth]{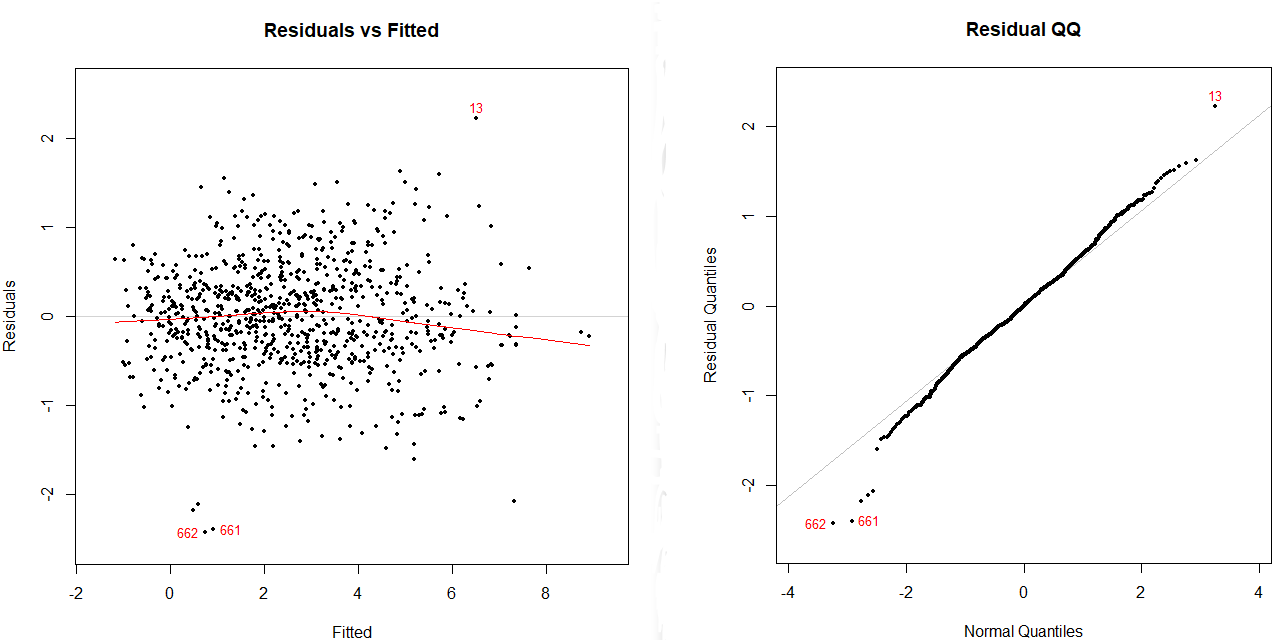}
	\caption{Residuals analysis of MARS model 2 }
	\label{fig:screenshot064}
\end{figure}

Figure \ref{fig:screenshot064} shows the residual plots of the second MARS model. It can be seen that the residuals vs fitted  plot shows no dependency in errors and they appear to have almost constant variance. Furthermore, the residuals QQ plot shows that the errors are normally distributed. 

\section{Artificial neural network model}\label{ANN}
The artificial neural network (ANN)   can learn from  data and detect  patterns and relationships in data  \citep{Agatonovic-Kustrin}.  ANNs can  be used to  predict the output from available information. The neural structure of an ANN  consists of a number of neurons which are known as processing elements \citep{Erzin},  where each processing element represents an equation that consists of weighted inputs, transfer function and one output \citep{Agatonovic-Kustrin}.

We explore the use of ANNs to predict the electrical resistivity of soil based on its geotechnical properties. Three ANN models (ANN.model1, ANN.model2 and ANN.model3) were developed for predicting ER. 

The first model,  ANN.model1, takes into consideration the effect of ST.KB, ST.KS, Mol, Moist and Uw, whereas the second model,  ANN.model2, takes into account the effect of ST.K, Mol, Moist and Uw.  The third model,  ANN.model3,  takes into account the effects that are in  ANN.model2 plus the effect of interaction between  ST.K and  Moist, and the interaction between  ST.K and Uw. All three models have the output ER. 

In most applications, a  single hidden layer is enough for reasonable approximation \citep{Agatonovic-Kustrin}. So, one hidden layer is chosen for each model. \citep{Erzin} pointed out that the upper limit for the number of neurons in the hidden layer is equal to $2i+1$ where $i$ is the number of input parameters. Hence, the number of neurons in the hidden layer  for ANN.model1, ANN.model2 and ANN.model3 should not be greater than 13, 11 and 15,  respectively. To get the optimum number of neurons in the hidden layer of each model, we  fitted each model with one neuron in the hidden layer and then increased the number of neurons  to the upper limit. Moreover, we randomly divided the data into a training dataset (75\% of the original data) and an independent validation dataset (25\% of the original data) to estimate  the performance of  the models.

\begin{figure}[h!]
	\centering
	\includegraphics[width=0.85\linewidth]{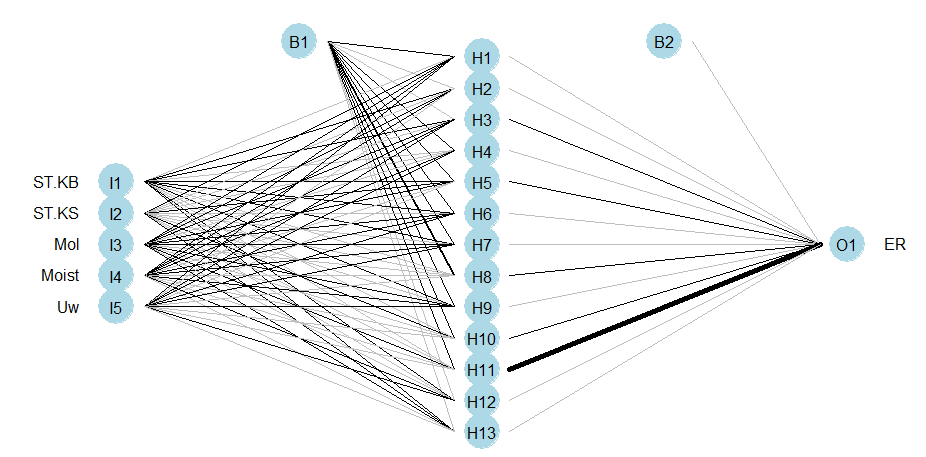}\vspace{1mm}
	\caption{Neural network plot for ANN.model1.}
	\label{fig:screenshot055}
\end{figure}

\begin{figure}[h!]
	\centering
	\includegraphics[width=0.85\linewidth]{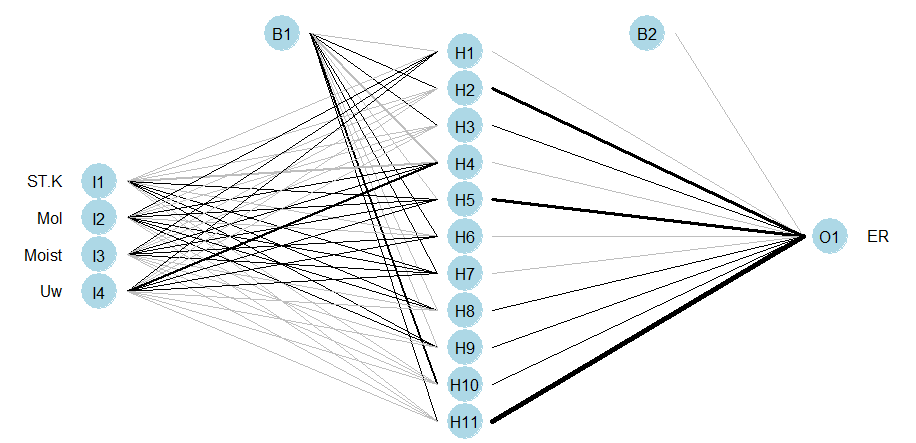}\vspace{1mm}
	\caption{Neural network plot for ANN.model2.}
	\label{fig:screenshot056}
\end{figure}

\begin{figure}[h!]
	\centering
	\includegraphics[width=0.85\linewidth]{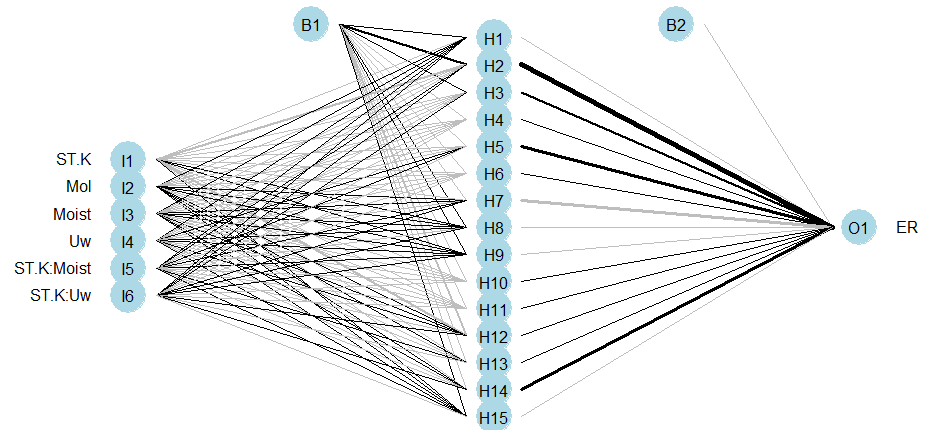}\vspace{1mm}
	\caption{Neural network plot for ANN.model3.}
	\label{fig:screenshot057}
\end{figure}

\begin{figure}[h!]
	\centering
	\includegraphics[width=01\linewidth,height=7cm]{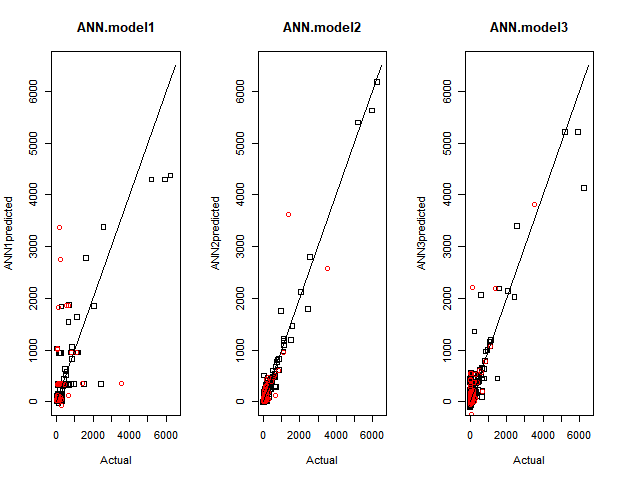}
	\caption{Predicted ER vs actual ER for the three ANN Models. }
	\label{fig:screenshot054}
\end{figure}

The neural network plots for each model are shown in Figure \ref{fig:screenshot055}, \ref{fig:screenshot056} and \ref{fig:screenshot057}. Note that each model has two bias layers (B1 and B2). 
\newpage
Figure~\ref{fig:screenshot054}  shows the predicted value of ER vs actual values for each model. It can be seen that most of points lie very close to the  reference lines, especially for ANN.model2.  The $\Box$ symbols represent the training sample and the $\circ$ symbols represent the validation sample. The MSE errors for ANN.model1, ANN.model2 and ANN.model3 are 182060.6, 49255.5 and 63702.84 respectively.

We can conclude that  ANN.model2 is  the best model and it can  be used to reliably predict the electrical resistivity  of soil.

\section{Comparison and validation of models}\label{S:Comparing Models}
We  explored various statistical models to fit the electrical resistivity of soil and its geotechnical properties. The methods that were used include linear,   non-linear regressions  and artificial neural networks (ANN). Each model has its own advantages and disadvantages. Unfortunately, there is not  single  statistic  that can be used  to choose  which one of these  models  is the best. Therefore, a particular statistic was used for each method. For example, the R-squared was used to compare linear models  fitted using OLS approaches, and MSE and AIC were used to compare non-linear models. 

To confirm our findings and compare all models we used the cross-validation analysis. The data were randomly divided into a training dataset (75\% of the original data) and a validation dataset (25\% of the original data). Then, the best four models, 
\begin{align*}
\log(\widehat{ER})= &\ 16.05 -5.23 ST.K  -0.90Mol  -0.246Moist
	 -6.30Uw\notag\\
	  &+\ 0.05ST.K\times Moist +  2.83 ST.K\times Uw;\notag\\
	 \widehat{ER}= &\ 57.62 +145546403\cdot 36.67^{-ST.K}1.20^{-Moist}11275.04^{-Uw}-32.97Mol;\notag\\
	\widehat{ER} =&\ \exp[4.958 -   0.8739  ST.K  +   2.808  \max(0,  0.5 -  Mol)     \notag\\ 
	& -    0.9058\max(0,   Mol -   0.5)+     0.739  \max(0,   Mol -     1)   \notag\\ 
	& -    0.2447  \max(0, Moist -     5) +   0.07649  \max(0, Moist -    15)  \notag \\
	&+     105.9  \max(0,    Uw - 1.005)  -       119  \max(0,    Uw - 1.014) \notag\\
	& +     12.28  \max(0,    Uw - 1.099)  +     2.504 \max(0, 1.663 -    Uw) ]; 
\end{align*}
and the artificial neural network model ANN.model2, which input layer consists of ST.K, Mol, Moist and  Uw, were refitted  using only the training dataset. The new fitted models were applied to the  validation dataset  to predict ER. MSE was calculated for each model. The procedure was repeated 100 times. Each time, new training and validation data sets were randomly selected. The mean of MSEs for each model was calculated, see Table~\ref{MSEs}. The result shows that the most accurate prediction values of the electrical resistivity were obtained using the ANN model.
\begin{table}[h!]
	\caption{MSEs for the best four models} 
	\label{MSEs}
	\centering
	\begin{tabular}{c|cccc}
		\hline\\[-1mm]
		& Lin	&  NonLin3 & MARS2  & ANN2  \\[1mm]
		\hline\\[-1mm]
		Mean MSE & 67827.3 & 52639.23  & 84044.25 & 38831.16 \\
		\hline
	\end{tabular}
\end{table}

Based on  all the all above evidence we conclude that the neural network model (ANN.model2) is the best model for predicting electrical resistivity. 

\section{Conclusion and future work}
The main objective of this study was to develop an accurate  statistical model that can be used to estimate the electrical resistivity of soil  based on the knowledge of  soil type (ST), molarity (Mol), moisture (Moist) and unit weight (Uw).

The obtained results showed that there is a significant exponential negative relationship between electrical resistivity and all  four predictors. Moreover, it was demonstrated that treating soil composition as a factor variable (denoted by ST.K) on the basis of pure kaolin or not pure kaolin is better than treating it as a continuous variable  based on the percentage of bentonite (ST.KB) and sand (ST.KS). The interactions between ST.K and moisture, and between ST.K and Uw were found to be significant predictors for first two models. As a result, both these interactions were included in the final models.   Moreover, the log transformation of the dependent variable yields better models. In particular, it substantially improved the value of R-squared in  the linear models from  about 0.20 to  more than 0.80.

The analysis was performed to identify the optimal combination of the degree of interactions and the number of retained terms or complexity of neural networks. For  MARS2 and ANN2 models it was shown that the interaction between ST.K and Uw does not contribute significantly to electrical resistivity prediction. A plausible explanation is that MARS2 and ANN2 models use more flexible non-linear transformations compare to the second model. They can reduce prediction errors directly via separately transformed ST.K and~Uw.  

Several useful   models  have been developed and the results  from each model were compared with the actual values of electrical resistivity using the cross-validation. In addition,  performance indices such as the coefficient of determination (R-squared), MSE and  AIC were used to assess the performance of the models. 

It was found that the ANN model is able to efficiently predict the electrical resistivity of soil and is better than the other models that were developed. 

The developed ANN model in this study could contribute to the continuous research effort in the field of electrical resistivity imaging as it captures the influence of different geotechnical properties of soil on its electrical resistivity. Furthermore, it sheds light on what has been learnt throughout the study in terms of the sensitivity of different soil variables and how its incremental changes in the experimental program should be chosen to maximize the accuracy of the developed model. In this regard, it is recommended that, in the future experimental studies, the incremental change in the percentage of bentonite and sand, salinity of pore water, moisture content, and dry unit weight should be smaller than the values used in this study. In fact, this recommendation can be justified by the observed exponential relationships between electrical resistivity of soil and its geotechnical properties which causes a large change in electrical resistivity even with a small change in geotechnical properties. However, experts in the field know that very small incremental changes in the geotechnical parameters of soil in laboratory experiments are often difficult to be achieved accurately. Therefore,  their effects on measurement results could be lower than the true ones. It is desirable to implement experimental designs that balance the size of increments and the number of repeated measurements. The problem of balancing and assessing accuracy of such tests is an interesting direction for future research.

For fully automated intelligent system analysis based on the first two models,  transformation of data is an area of potential improvement. The choice of transformation method is conventionally based on descriptive statistics results. This process can be semi-supervised if a finite number of admissible functional transformations and relationships is specified.

One of the main known ANN limitations is knowledge extraction from trained ANNs, i.e. interpreting ANN models similar to the other methods that provide functional relation to input effects. The Lek's profile and local interpretable model-agnostic explanations  methods, see \citep{Zhang2}, could be potentially useful approaches in interpreting ANN results. 

For small randomly selected test or training sets ANN resulting models can substantially vary, nevertheless, providing rather similar predictions. Studies with a large number of explanatory parameters can also result in a large set of candidate models. In such studies it would be interesting to quantitatively investigate agreement levels between top selected models by using various agreement coefficients, see \citep{Olenko}.

Finally, in the future research it is important to investigate other parameters like the proctor density of the soil, ambient temperature,  two-layer or a multi-layer stratification to obtain finer electro resistivity models.

\section*{Acknowledgement}
The authors are grateful to the referees for their careful reading of the paper and suggestions that helped to improve the manuscript.


\begin{thebibliography}{99}
	\bibitem[Abu-Hassanein et al.(1996)]{Abu-Hassanein} Abu-Hassanein, Z. S., Benson, C. H., \& Blotz, L. R. (1996). Electrical resistivity of compacted clays. \textit{Journal of Geotechnical Engineering,} 122(5), 397-406.
	
	\bibitem[Ackerman et al.(2013)]{Ackerman}  Ackerman, A.,  Sen, P. K., \&  Oertli, C. (2013). Designing safe and reliable grounding in AC substations with poor soil resistivity: An interpretation of IEEE Std. 80.  \textit{IEEE Transactions on Industry Applications,}  49(4), 1883-1889. 
	
	\bibitem[Agatonovic-Kustrin et al.(2000)]{Agatonovic-Kustrin} Agatonovic-Kustrin, S.,  \& Beresford, R. (2000). Basic concepts of artificial neural network  modeling and its application in pharmaceutical research. \textit{Journal of Pharmaceutical and Biomedical Analysis,} 22(5), 717-727. 
	
	\bibitem[Al Rashid et al.(2018)]{Al Rashid} Al Rashid, Q.A., Abuel-Naga, H.M., Leong,  E.-C.,  Lu, Y., \& Al~Abadi,~H.~(2018).  Experimental-artificial intelligence approach for characterizing electrical resistivity of partially saturated clay liners. \textit{Applied Clay Science,} 156, 1-10.
	
	\bibitem[Alipio \&   Visacro(2013)]{Alipio}  Alipio, R. \&  Visacro, S.  (2013) Frequency dependence of soil parameters: effect on the lightning response of grounding electrodes. \textit{IEEE Transactions on Electromagnetic Compatibility}, 55(1), 132-139.
	
	\bibitem[Alipio \&   Visacro(2014)]{Alipio2}  Alipio, R. \&  Visacro, S.  (2014). Modeling the frequency dependence of electrical parameters of soil. \textit{IEEE Transactions on Electromagnetic Compatibility}, 56(5), 1163-1171.
	
	\bibitem[ASTM G57(2006)]{ASTM}	ASTM G57 (2006). {\it Standard Test Method for Field Measurement of Soil Resistivity Using the Wenner Four-Electrode Method}. West Conshohocken: ASTM International.
	
	\bibitem[Brillante et al.(2015)]{Brillante} Brillante, L., Mathieu, O., Bois, B., van Leeuwen, C., \&  L\'{e}v\^{e}que, J. (2015). The use of soil earthing system performance. \textit{Rev. Energ. Ren.: Power Engineering}, 57-61.
	
	\bibitem[Clavel et al.(2018)]{Clavel}  Clavel, E.,  Roudet, J.,  Guichon, J. M.,  Gouichiche, Z.,  Joyeux,  P., \&   Derbey, A. (2018). A nonmeshing approach for modeling grounding.  \textit{IEEE Transactions on Electromagnetic Compatibility,} 60(3), 795-802.
	
	\bibitem[Datsios et al.(2017)]{Datsios} Datsios, Z. G.,  Mikropoulos, P. N., \&  Karakousis, I, (2017). Laboratory characterization and modeling of DC electrical resistivity of sandy soil with variable water resistivity and content. \textit{IEEE Transactions on Dielectrics and Electrical Insulation},  24(5),  3063-3072.
	
	%\bibitem{Davidian} Davidian, M., \& Giltinan, D. (1998). \textit{Nonlinear Models for Repeated Measurement Data}. Boca Raton, Fla.: Chapman \& Hall/CRC.
	
	\bibitem[Edlebeck \&  Beske(2014)]{Edlebeck} Edlebeck, J. E., \&  Beske, B.  (2014) Identifying and quantifying material properties that impact aggregate resistivity of electrical substation surface material. \textit{IEEE Transactions on Power Delivery},  29(5),  2248-2253.
	
	\bibitem[Erzin et al.(2010)]{Erzin} Erzin, Y.,  Rao, B., Patel, A.,  Gumaste, S.,  \& Singh,~D. (2010). Artificial neural network models for predicting electrical resistivity of soils from their thermal resistivity. \textit{International Journal of Thermal Sciences,} 49(1), 118-130. % doi:10.1016/j.ijthermalsci.2009.06.008.
	
	\bibitem[Fox \&  Weisberg(2010)]{Fox} Fox, J., \&  Weisberg, S. (2010). \textit{Nonlinear Regression and Nonlinear Least Squares in R  Appendix to an R Companion to Applied Regression.}   \url{http://socserv.socsci.mcmaster.ca/jfox/Books/Companion/appendix/Appendix-Nonlinear-Regression.pdf} Accessed 21 July 2018.
	
	\bibitem[Friedman(1991)]{Friedman} Friedman, J, (1991). Multivariate adaptive regression splines. \textit{The Annals of Statistics,} 19(1), 1-67. %Retrieved from\\ http://www.jstor.org/stable/2241837
	
	\bibitem[Gomes et al.(2018)]{Gomes}  Gomes, T. V., Schroeder, M. A. O.,  Alipio, R.,  de Lima, A. C. S., \&  Piantini, A. (2018). Investigation of overvoltages in hv underground sections caused by direct strokes considering the frequency-dependent characteristics of grounding, \textit{IEEE Transactions on Electromagnetic Compatibility}, to appear.
	
	\bibitem[Gen\c{c}o\u{g}lu,  M.T. \& Cebeci(2009)]{Gen} Gen\c{c}o\u{g}lu,  M.T. \& Cebeci, M. (2009). Investigation of pollution flashover on high voltage insulators using artificial neural network, \textit{Expert Systems with Applications}, 36(4), 7338-7345.

	\bibitem[Griffiths \& King(1965)]{Griffiths} Griffiths, D. H. \& King, R. F. (1965). \textit{Applied Geophysics for Engineers and Geologists}. Oxford: Pergamon Press.
	
	\bibitem[Gusel \& Brezocnik(2011)]{Gusel} Gusel, L. \& Brezocnik, M. (2011). Application of genetic programming for modelling of material characteristics, \textit{Expert Systems with Applications}, 38(12), 15014-15019. 
	
	\bibitem[Hastie et al.(2009)]{Hastie} Hastie, T., Tibshirani, R., \& Friedman, J. (2009). \textit{The Elements of Statistical Learning: Data Mining, Inference and Prediction}. New York: Springer-Verlag.
	
	\bibitem[Hwang et al.(2010)]{Hwang} Hwang, R.-C.  Chen, Yu-Ju, \&  Huang,  H.-C. (2010). Artificial intelligent analyzer for mechanical properties of rolled steel bar by using neural networks, \textit{Expert Systems with Applications}, 37(4), 3136-3139.
	
	\bibitem[Jones et al.(2011)]{Jones} Jones, C.J.F.P. Lamont-Black, \& J. Glendinning S. (2011). Electrokinetic geosynthetics in hydraulic applications, \textit{Geotextiles and Geomembranes,} 29, \mbox{381-390.}
	
	\bibitem[Kibria \& Hossain(2012)]{Kibria} Kibria, G., \& Hossain, M. S. (2012). Investigation of geotechnical parameters affecting electrical resistivity of compacted clays. \textit{Journal of Geotechnical and Geoenvironmental Engineering,} 138(12), 1520-1529.
	
	\bibitem[Laver \& Griffiths(2001)]{Laver} Laver, J. A., \& Griffiths, H. (2001). The variability of soils in earthing measurements and earthing system performance. \textit{Rev. Energy Ren. Power Engineering, Cardiff University, UK,}~57-61.
	
	\bibitem[Lim et al.(2013)]{Lim} Lim, S. C, Gomes, C., \& Kadir, M. Z. (2013). Earthing in troubled environment. \textit{International Journal of Electrical Power \& Energy Systems,} 47, 117-128.
	
	\bibitem[Mokhtari \&  Gharehpetian(2018)]{Mokhtari}  Mokhtari, M. \&  Gharehpetian, G. (2018). Integration of energy balance of soil ionization in CIGRE grounding electrode resistance model. \textit{IEEE Transactions on Electromagnetic Compatibility,} 60(2), 402-413. 
	
	\bibitem[Montgomery et al.(2012)]{Montgomery} Montgomery, D. C., Peck, E. A., \& Vining, G. G. (2012). \textit{Introduction to Linear Regression Analysis}. Hoboken, NJ: \mbox{Wiley.}
	
	\bibitem[Olenko \&  Tsyganok (2016)]{Olenko} Olenko, A., \& Tsyganok, V. (2016). Double entropy inter-rater agreement indices. \textit{Applied Psychological Measurement,} 40(1), 37-55.
	
	\bibitem[Rhoades et al.(1976)]{Rhoades} Rhoades, J., Raats P.A.C, \& Prather, R. (1976). Effect of liquid-phase electrical conductivity, water content and surface conductivity on bulk soil electrical conductivity. \textit{Soil Sci. Soc. of Am. J.,} 40, 651-655. 
	
	\bibitem[Sakia(1992)]{Sakia} Sakia, R. (1992). The Box-Cox transformation technique: A Review.\textit{ Journal of the Royal Statistical Society,} 41(2), 169-178. %doi:10.2307/2348250.
	
	%\bibitem{Shanmuganathan} Shanmuganathan, S., \& Samarasinghe, S. (2016).\textit{ Artificial Neural Network Modelling}. New York: Springer.% Springer Publishing Company, Incorporated.
	
	\bibitem[Southey et al.(2015)]{Southey} Southey,  R. D., Siahrang,  M., Fortin,  S., \&  Dawalibi, F. P. (2015). Using fall-of-potential measurements to improve deep soil resistivity estimates. \textit{IEEE Transactions on Industry Applications,}  51(6), 5023-5029.
	
	\bibitem[Tabbagh et al.(2002)]{Tabbagh} Tabbagh, A., Benderitter, Y., Michot, D., \& Panissod, C. (2002). Measurement of variations in soil electrical resistivity for assessing he volume affected by plant water uptake. \textit{European Journal of Environmental and Engineering Geophysics}~7, 229-237.
	
	%\bibitem{Tino} Tino, P., Benuskova, L., \& Sperduti, A. (2015). Artificial neural network models. In J. Kacprzyk \& W. Pedrycz  (Eds). \textit{Springer Handbook of Computational Intelligence}, 455-471. %doi:10.1007/978-3-662-43505-2\_27
	
	\bibitem[Wenner(1915)]{Wenner}	Wenner, F. (1915) A method for measuring earth resistivity. \textit{Bulletin of the Bureau of Standards,} 12, 469-478.
	
	\bibitem[Zastrow(1979)]{Zastrow} Zastrow, O. W., (1979). Choices of jacketed or bare concentric neutral cable for effective grounding and corrosion control. \textit{IEEE Transactions on Industry Applications}, IA-15(1), 80-84.
	
	\bibitem[Zhang \& Goh(2016)]{Zhang} Zhang, W., \& Goh, A. T. (2016). Multivariate adaptive regression splines and neural network models for prediction of pile drivability. \textit{Geoscience Frontiers,} 7(1), 45-52.% doi:10.1016/j.gsf.2014.10.003
	
	\bibitem[Zhang et al.(2018)]{Zhang2} Zhang, Z., Beck, M. W., Winkler, D. A., Huang, B., Sibanda, W.,  \& Goyal,~H. (2018). Opening the black box of neural networks: methods for interpreting neural network models in clinical applications. \textit{Annals of translational medicine,} 6(11): 216. 
	
	\bibitem[Zhou Wang et al.(2015)]{Zhou} Zhou Wang, J.,  Cai, L.,  Fan, Y., \&  Zheng, Z., (2015). Laboratory investigations on factors affecting soil electrical resistivity and the measurement. \textit{IEEE Transactions on Industry Applications,}  51(6), \mbox{5358-5365.}
\end{thebibliography}
\end{document}